\documentclass[pre,amsmath,twocolumn,showpacs,superscriptaddress]{revtex4}

\usepackage{amssymb}
\usepackage{epsfig}
\usepackage{graphics}
\usepackage{natbib}

\newcommand{\xdist}{x_{\rm dist}}
\newcommand{\fdist}{f_{\rm dist}}
\newcommand{\etadist}{\eta_{\rm dist}}
\newcommand{\etainit}{{\bar\eta}^{\rm init}}
\newcommand{\noise}{{\bar \eta}^{\rm eff}}
\newcommand{\wid}{b}
\newcommand{\tdist}{{\tau_{\rm dist}}}
\newcommand{\mT}{{\mathcal{T}}}
\newcommand{\kdist}{k_{\rm dist}}
\newcommand{\nd}{{n_{\rm d}}}
\newcommand{\na}{{n_{\rm a}}}
\newcommand{\fd}{{n_{\rm d}}}
\newcommand{\fa}{{n_{\rm a}}}
\newcommand{\aspr}{\simeq}
\newcommand{\aseq}{\sim}
\newcommand{\msd}{\langle \delta x_\mT^2(t)\rangle}
\newcommand{\Sf}{S_0}

\begin{document}

\title{Foundation of Fractional Langevin Equation: Harmonization of a Many Body Problem}
\author{Ludvig Lizana}
\affiliation{Niels Bohr Institute, Blegdamsvej 17, DK-2100, Copenhagen, Denmark}                                                                             
\author{Tobias Ambj\"ornsson}
\affiliation{Department of Theoretical Physics, Lund University, SE-223 62 Lund, Sweden}

\author{Alessandro Taloni}
\affiliation{School of Chemistry, Tel Aviv University, Tel Aviv 69978, Israel}

\author{Eli Barkai}
\affiliation{Department of Physics, Institute of Nanotechnology and Advanced Materials, Bar Ilan University, Ramat-Gan 52900 Israel}

\author{Michael A. Lomholt}
\affiliation{MEMPHYS, University of Southern Denmark, Campusvej 55, 5230 Odense M, Denmark}

\pacs{05.40.-a, 02.50.Ey, 82.37.-j}

%
%

\begin{abstract}
In this study we derive a single-particle equation of motion, from first-principles, starting out with a microscopic description of a tracer particle in a one-dimensional many-particle system with a general two-body interaction potential. Using a new {\it harmonization} technique, we show that the resulting dynamical equation belongs to the class of fractional Langevin equations, a stochastic framework which has been proposed in a large body of works as a means of describing anomalous dynamics. Our work sheds light on the fundamental assumptions of these phenomenological models.
\end{abstract}

\maketitle

%
%

\section{Introduction}
The stochastic dynamics of many-body systems with general two-body interactions are inherently difficult to solve. There are, however, a few idealized exactly solvable models that have served as benchmark cases from which collective effects have been deduced \cite{derrida1998exactly,mattis1993many, lebowitz1966mean}.  One example is the one-dimensional over-damped motion of non-passing hard spheres (so called single-file diffusion) in which a tracer particle  behaves subdiffusively \cite{percus1974anomalous,gupta2007tagged}. Processes displaying anomalous diffusion occurs in a range of many-body systems \cite{ben2009strong,buttiker1980long,imry1974correlation}, especially in biology \cite{banks2005anomalous,golding2006physical}, and are, apart from a few exceptional cases, modeled in phenomenological ways.

The one-dimensional motion of identical Brownian particles (BPs) which are unable to pass each other is well studied theoretically
\cite{harris1965diffusion,levitt1973dynamics,van1983diffusion,kollmann2003single, lizana2008single}. A tracer particle in such single-file system exhibits sub-diffusion; its mean square displacement (MSD) is proportional to $t^{1/2}$ indicating slow dynamics \cite{harris1965diffusion} \cite{note1}.
There exists a wide range of experiments on tracer particle dynamics in diverse systems which show the $t^{1/2}$-behavior. Examples include  colloids in one-dimensional channels \cite{lutz2004single, wei2000single, lin2005random}, an NMR experiment involving Xenon in microporous materials \cite{meersmann2000exploring}, molecular diffusion in Zeolites \cite{hahn1996single, karger1992diffusion}, moisture expansion in ceramic materials \cite{wilson2003kinetics}, and a study of Ethane in a molecular sieve \cite{gupta1995evidence}.

Independent of the developments in the field of single-file diffusion, recently a stochastic fractional Langevin equation (FLE) has gained much interest  \cite{lutz2001fractional,kou2004generalized, burov2008critical}.
In the FLE a derivative of fractional order $\alpha$ replaces the usual first order time derivative in the overdamped Langevin equation ($d/dt \to d^\alpha/dt^{\alpha}$).
FLEs with 0$<$$\alpha$$<$1 are able to describe phenomenologically a range of physical phenomena such as motion in viscoelastic media \cite{goychuk2009viscoelastic}.
In the presence of a binding harmonic field, Xie and co-authors used the FLE to model protein dynamics \cite{kou2004generalized,min2005observation} and  $\alpha=0.51\pm0.07$ was deduced from experimental observations. Here we derive an FLE with $\alpha = 1/2$, in the presence of an external force field, starting from a many-body theory. Thus we show that the 1/2-FLE is expected to be universal for a large number of experiments describing interacting tracer particle dynamics.

Usually single-file models consider BPs with hard-core interaction. These models
can be mapped onto a non-interacting system with known methods 
\cite{levitt1973dynamics,barkai2009theory}. In experiments the interaction of BPs is
hardly ever hard-core. In this paper we consider rather
general two-body interactions between BPs. We present a new method
to deal with this many-body problem, which we call harmonization.
With this method we are able to effortlessly derive many previous results, for
example the Kollmann relation for the MSD \cite{kollmann2003single},
to justify from first-principles the FLE,
and to derive many new results, such as the distance correlation
function between a pair of particles.

%
%

\begin{figure}
\begin{center}
\includegraphics[width=0.45\textwidth]{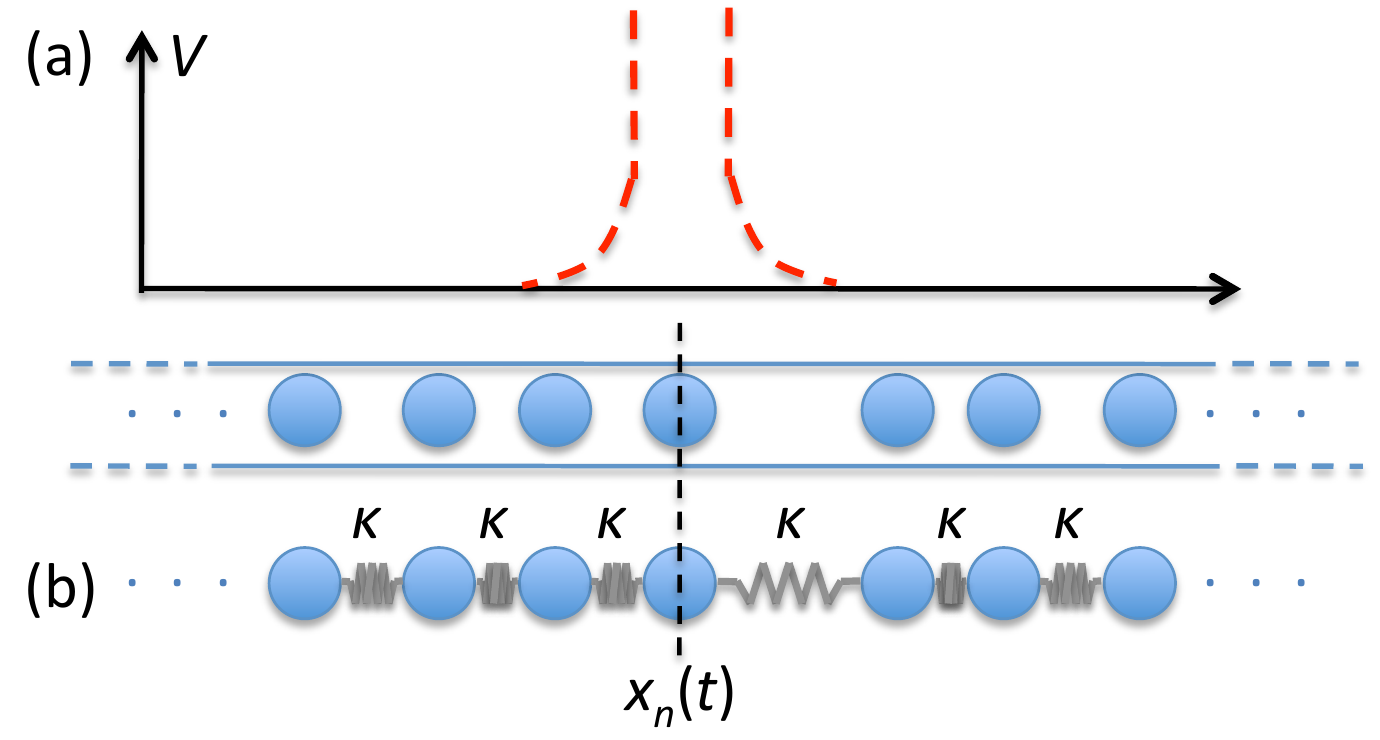}
\end{center}
\caption{Illustration of the many-particle system studied here. (a) Particles interact via the two-body potential ${\cal V}(|x_n(t) - x_{n'}(t)|)$ and cannot pass each other. (b) The harmonization procedure amounts to mapping the interacting particle system onto a harmonic chain in which the interactions are captured by the effective spring constant $\kappa$.}
\label{figSchematic}
\end{figure}

\section{System description}
We consider particles undergoing one-dimensional over-damped Brownian motion
in an infinite system where two particles $n$ and $n'$ interact via the
two-body potential ${\cal V}(|x_n(t) - x_{n'}(t)|)$, where $x_n(t)$ is the
position of the $n$th particle. The potential has a hard-core part, so
the particles cannot pass each other,
but otherwise it is rather general
\cite{note2}.
The Langevin equation for the motion of particle $n$ reads
\begin{equation}
\xi \frac{dx_n(t)}{dt}  = \sum_{n'}  \mathfrak{f}[x_n(t) - x_{n'}(t)]
+\eta_n (t) + f_n (t), 
\label{eq01}
\end{equation}
where $\xi$ is the friction constant ($D=k_BT/\xi$ is the free-particle
diffusion coefficient), $\mathfrak{f}=-\partial {\cal V}/ \partial x_n$ 
is the force due to interactions with the surrounding particles,
$\eta_n$ is a zero mean white Gaussian noise with $\langle \eta_n(t)
\eta_{n'}(t') \rangle = 2 \xi k_B T \delta(t - t')\delta_{n n'}$ where $k_B$
and $T$ are the Boltzmann constant and temperature.  $f_n$ is an
external force. The particles are initially taken to be in thermal equilibrium.
Our main interests are in the dynamics of a tracer particle position and in
distance fluctuations between particles. These quantities are, however,
intractable from the many-body problem (\ref{eq01}) for a general ${\cal
V}$. Therefore, we introduce a new technique
- harmonization.

%
%
\section{Harmonization technique}
The philosophy behind harmonization is to map the original system (A), i.e., a
system described by Eq. (\ref{eq01}), to a system (B) consisting of beads
interconnected by harmonic springs. Consider two particles in system A with
$N\gg 1$ particles in between, and which, in equilibrium, are at an average
distance $L_{\rm eq}$ from one another. We let $\mathcal{F}(L,N)$ be the extensive free energy due to the $N$ particles at a fixed temperature $T$.
Fluctuations of $L$ around $L_{\rm eq}$ are small. Hence
we may expand $\mathcal{F}(L,N)\sim \mathcal{F}(L_{\rm eq},N) +
k_N (L- L_{\rm eq})^2/2$
where we introduced a macroscopic spring constant:
$k_N=\partial^2 \mathcal{F}(N,L) / \partial L^2|_{L_{\rm eq}}$. Now we replace system A, by a
system B of beads connected to their two nearest neighbours by springs with
spring constant $\kappa$. We relate the spring constants $\kappa$ and $k_N$
by requiring that the total free energies associated with identical
displacements from
$L_{\rm eq}$ in the two systems are the same. This gives
\begin{equation}
\kappa = N k_N = N \left.\frac{\partial^2 \mathcal{F} (N,L)}{ \partial L^2}\right|_{L_{\rm eq}}.
\label{eq02}
\end{equation}
To see this note that for $N$ particles in system B the total free energy
change is $N \kappa ( L-L_{\rm eq})^2/(2 N^2)$ [since the displacement of one
spring is $(L-L_{\rm eq})/N$ and we have $N$ such springs] which is equal to
the free energy of a single macroscopic spring in system A, $k_N (L-L_{\rm
eq})^2/2$, when Eq. (\ref{eq02}) holds. The effective spring constant
$\kappa$ is an intensive thermodynamic quantity, which can be obtained from the original
system's equation of state. From the definition of pressure $P =- (\partial
{\cal F} / \partial L)_N$ and isothermal compressibility $\chi_T= -
(\partial L/ \partial P)_N/L$,
one obtains $\kappa =- N (\partial P/ \partial
L)_N = \rho/\chi_T$ where $\rho=N/L$ ($N,L \rightarrow \infty$) is the
particle number density. For a one-dimensional gas of hard-core
interacting point-particles the equation of state is given by
$P = N k_B T/L$  which leads to $\kappa = \rho^2 k_B T$. Similarly, for systems consisting of $\wid$-sized hard rods (Tonks gas) we have
from the van der Waals equation $P = Nk_B T/(L - N\wid)$ that
$\kappa = \rho^2 k_B T (1 - \rho \wid)^{-2}$.

The final step in the harmonization method is to replace the non-linear
two-body interaction in Eq. (\ref{eq01}) by forces from the
nearest-neighbor spring coupling, i.e.,
\begin{equation}
\xi \frac{dx_n(t)}{dt} = \kappa 
	\left[ 
			x_{n+1}(t)+ x_{n-1}(t) - 2 x_n(t)
	\right]
	+ f_n(t) + \eta_n(t).
\label{eq03}
\end{equation} 
We above implicitly assumed that there has been time for particles to interact with neighbors, $t\gg \tau_{\rm int}=1/(\rho^2D)$. Equation (\ref{eq03}) will be
justified later, via simulations, and below by showing agreement with known
results for single-file systems.
Under the assumption $t\gg \tau_{\rm int}$
we can take the continuum limit and turn $x_n(t)$ into a field
\begin{equation}
\xi \frac{\partial x(n,t) }{ \partial t} = \kappa \frac{ \partial^2 x(n,t) }{ \partial n^2} + f(n,t) +  \eta(n,t).
\label{eq04}
\end{equation}
This relationship is our harmonization equation from which previous exact
results are recovered and new ones derived. In the following subsection the MSD of a tracer
particle is discussed. We note that a mapping similar to Harmonization was applied to the simple exclusion process in \cite{gupta2007tagged}. Here we consider a general two body interaction showing precisely how to compute the effective spring constant from equilibrium concepts (i.e., the compressibility).

%
%

\subsection{Mean square displacement}
Consider the case of no external forces,
$f(n,t)=0$. Equation (\ref{eq04}) is then equivalent to the Rouse model from polymer physics (see for instance \cite{grosberg1994statistical}). In the following we consider a
tracer particle
labeled $\mT$ and consider its MSD. We will arbitrarily choose the particle $n=0$ with position $x_\mT(t)=x(n=0,t)$. The MSD, $\msd=\langle [x_\mT(t)-x_\mT(0)]^2\rangle$ ($\langle\,\rangle$
denotes average over noise and initial conditions), is \cite{note3}
\begin{equation} 
\msd = k_BT \sqrt{{4t}/({\pi \xi\kappa }) }.
\label{eq05}
\end{equation}
The derivation is relegated to Appendix \ref{app:msd}.
If $\kappa$ for the gas of $b$-sized hard rods is used one obtains
$\msd = \rho^{-1}(1 -\rho \wid)\sqrt{ 4Dt/\pi }$,
which agrees with
\cite{ percus1974anomalous,alexander1978diffusion, fedders1978two, gonalves2008scaling,lizana2008single}. 
In Ref. \cite{kollmann2003single} Kollmann showed that $\msd \aspr t^{1/2}$
regardless of the nature of interactions as long as mutual passage is
excluded. In particular Kollmann derived the relation
$\msd \aseq \Sf \sqrt{4D_c t/(\rho^2\pi) }$, where
$\Sf$ is the static structure factor at zero wavevector and $D_c$ is the
collective diffusion constant.
Equation (\ref{eq05}) gives via Eq. (\ref{eq02}) a relation between the MSD and the free energy of the system, while the Kollmann relation relates the MSD to physical observables $\Sf$ and $D_c$.
Equivalence
between our results and \cite{kollmann2003single} is found using $\Sf=k_B T \rho
\chi_T$ and the relation $D_c=1/(\rho\chi_T \xi)$
\cite{note4}.

%
%

\section{Fractional Langevin equation}
Using our harmonization technique we recovered known single-file
results. Now we take the harmonization one step further and derive an
FLE for the position of the tracer particle $x_\mT(t)=x(n=0,t)$.
Taking the Fourier
and Laplace transforms \cite{note5}
$x(q,s)=\int_{-\infty}^\infty d n\int_0^\infty
d t\,e^{-i q n-s t} x(n,t)$ of Eq. (\ref{eq04}) gives 
\begin{equation}
x(q,s)=\frac{\eta(q,s)+\xi x(q,t=0)+f(q,s)}{\xi s+\kappa q^2}.
\label{eq10}
\end{equation}
Note that we for all functions indicate a Fourier transform with the variable $q$ and a Laplace transform by the variable $s$.
Subtracting $2\pi\delta(q)x_\mT(t=0)/s$ from both sides of (\ref{eq10}), rearranging, and
taking the inverse Fourier transform at $n=0$ yields
\begin{equation}
\gamma(s) [s x_\mT(s)-x_\mT(t=0)]=\noise_\mT(s)+{\bar f}_\mT(s),\label{eqLaplaceFLE}
\end{equation}
where $\gamma(s) = \sqrt{4\xi\kappa/s}$ is a fractional friction kernel, and the bar over a quantity means
${\bar y}_\mT(s)=\int_{-\infty}^\infty d n\,\exp({-\sqrt{\xi s/\kappa}|n|})y(n,s).$ 
The effective noise is defined as $\eta^{\rm eff}(n,s)=\eta(n,s)+\xi
[x(n,t=0)-x_\mT(t=0)]$ and includes Gaussian noise as well as randomness in
the initial conditions relative to the tracer particle. If the external
force acts only on the tracer particle, $f(n,t)=f_\mT(t)\delta(n)$, the inverse Laplace transform of Eq. (\ref{eqLaplaceFLE}) yields
\begin{equation}
\sqrt{4\xi \kappa}\frac{d^{1/2}x_\mT(t)}{d t^{1/2}}=\noise_\mT(t)+f_\mT(t),
\label{eqFLE}
\end{equation}
where we introduced the Caputo fractional derivative
\begin{equation}
\frac{d^{\alpha} f(t)}{d t^{\alpha}}=\frac{1}{\Gamma(1-\alpha)}\int_0^t\frac{dt'}{|t-t'|^{\alpha}}\frac{d f(t')}{d t'}
\end{equation}
of order $\alpha=1/2$.
Equation (\ref{eqFLE}) is the sought for FLE with fractional kernel
$\gamma(t)=\sqrt{4\xi \kappa/(\pi t)}$
\cite{note6}
and is in agreement with the long time
limit of the result proposed phenomenologically for hard-core interacting point particles in \cite{taloni2008langevin}. 
Notice that the
form $\gamma(t)\propto t^{-1/2}$
is a direct consequence of the harmonic
expansion and the assumption of over-damped dynamics.
Assuming thermal initial conditions one straightforwardly (see Appendix \ref{app:fle}) shows that the effective
noise satisfies the fluctuation-dissipation relation
\begin{equation}\label{eqFDT}
\left<\noise_\mT(t)\noise_\mT(t')\right>=k_BT\gamma(|t-t'|).
\end{equation}
This was expected since the Langevin equation was organized with
the external force conjugate to $x_\mT(t)$ as a term on the right hand side
(see \cite{kubo1966fluctuation}). For a constant force, $f_\mT(t)=F$, one can deduce
from Eq. (\ref{eqFLE}) (see Appendix \ref{app:force}) the 
generalized Einstein relation
\begin{equation}
\langle x_\mT(t) \rangle_F=F \msd/(2k_BT),\label{eq:einstrel}
\end{equation}
where $\langle x_\mT(t)\rangle_F$ is the average shift in position in the presence of the force and $\msd$ is the MSD in the absence of the force, i.e., as given by Eq. (\ref{eq05}); this
Einstein relation generalizes the results in \cite{burlatsky1996motion} to systems with general ${\cal V}$. For the case of a periodic force $f_\mT(t)=F_0\cos(\omega_0t)$,
we find asymptotically at long times a non-trivial $45^\circ$ phase-shift between the applied force and mean displacement
 (see Appendix \ref{app:force}):
\begin{equation}
\langle x_\mT(t)\rangle_{F_0} \sim \frac{F_0}{2\sqrt{\omega_0\kappa \xi}} \cos\left( \omega_0 t-\pi/4\right).\label{eq:perforce}
\end{equation}
The response of a tagged single-file particle to a harmonically oscillating force was previously obtained in a different way in \cite{taloni2008langevin}. More generally it has been obtained from the starting point of the fractional Langevin equation in \cite{burov2008pre}.

%
%

\begin{figure}
\begin{center}
\includegraphics[width=0.45\textwidth]{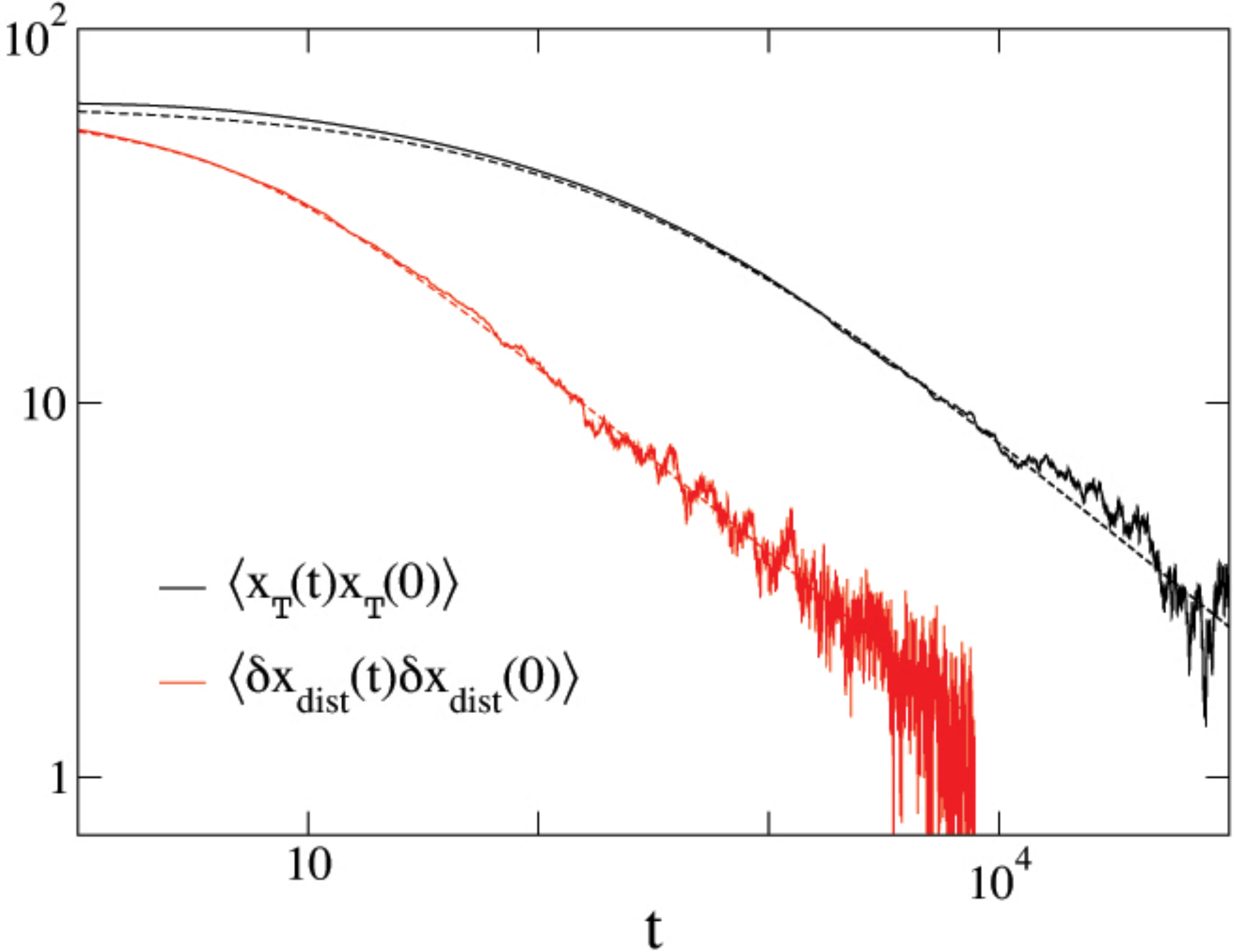}
\end{center}
\caption{Autocorrelation function obtained from molecular dynamics simulations of $10^4$ point particles on a ring (solid lines) with corresponding theoretical predictions (dashed lines). Upper black lines: tracer particle in an external harmonic potential with the Mittag-Leffler function
(\ref{eqPOTpred}) (parameters are: $\rho=0.25$, $k_BT=1$, $\xi=0.5$, $m\omega^2=0.0158$, resulting in $\tau \approx 501$).
Lower red lines: inter-particle distance correlation function with the prediction of Eq. (\ref{eqDISTpred}) (parameters are: $\rho=0.4$, $k_B T=1$, $\xi=0.5$, $\na-\nd=10$, giving $\tdist \approx 313$).
}
\label{fig1}
\end{figure}

\subsection{Tracer particle in a harmonic potential and simulations}
One of the predictions of FLE theory is that the autocorrelation function
$\langle x_\mT(t)x_\mT(0) \rangle$, under the influence of an external
harmonic potential, decays as a Mittag-Leffler function \cite{lutz2001fractional, kou2004generalized}. For the tracer particle in thermal equilibrium with respect to the harmonic force $f_\mT=-m\omega^2 x_\mT(t)$ we find from Eq. (\ref{eqFLE})
\begin{equation}
\langle x_\mT(t) x_\mT(0) \rangle = \frac{k_B T}{m\omega^2}E_{1/2}(-(t/\tau)^{1/2}),
\label{eqPOTpred}
\end{equation}
where
$E_{1/2}(-(t/\tau)^{1/2})=e^{t/\tau}{\rm erfc}[(t/\tau)^{1/2}]$
is the Mittag-Leffler function and $\tau=4\xi \kappa/(m\omega^2)^2$. For $t\gg \tau$ this leads to a decay $\langle x_\mT(t) x_\mT(0) \rangle\sim k_B T\sqrt{\tau/(\pi t)}/(m\omega^2)$. 
We tested the prediction of Eq. (\ref{eqPOTpred}) numerically by
simulations of hard-core interacting point particles. The result is
shown in Fig. \ref{fig1}; agreement with the analytic prediction is
excellent (without any fitting) in the time regime where particles have collided and harmonization is valid, i.e., when $t\gg \tau_{\rm int}=1/(\rho^2 D)$. The slight deviation at shorter times is in accordance with the interaction-free Brownian motion of the simulated particles prior to collisions, see also \cite{taloni2008langevin}.
An autocorrelation function with Mittag-Leffler decay of index 1/2 was recorded in the experiments
\cite{min2005observation}.
The harmonization procedure can be applied to other problems than tracer particles. This is illustrated below.

%
%
\section{Inter-particle distance correlations}
Donor-acceptor data from conformational dynamics of proteins \cite{yang2003protein} was
recently modeled using an FLE with a harmonic potential [i.e.,
Eq. (\ref{eqFLE}) with $f_\mT$ harmonic] \cite{kou2004generalized}. Fracton models
\cite{granek2005fractons} and FLEs based on the Kac-Zwanzig model \cite{kupferman2004fractional} have also recently been studied. Contrasting more phenomenologically oriented approaches,
our harmonization technique allows us to attack problems related to
inter-particle dynamics on a first-principle level. In fact, considering
inter-particle distance dynamics we will now show that: (i) a harmonic potential arises naturally
and is not due to an external field as assumed so far, 
and (ii)
the governing equation is a generalized Langevin equation (GLE) with
a power law memory kernel which leads to anomalous relaxation,
rather than an FLE.

Defining $\xdist(t)=x(\na,t)-x(\nd,t)$ as the distance between an ``acceptor'' particle $\na$ and ``donor'' $\nd$, an equation for $\xdist(t)$ is obtained by subtracting
Eq. (\ref{eqLaplaceFLE}) for $x(\nd,t)$ from the corresponding one for
$x(\na,t)$.  If external forces acting only on particles $\na$ and $\nd$
are considered,
$f(n,t)=f_\fa(t)\delta(n-\na) + f_\fd(t)\delta(n-\nd)$, we find 
\begin{eqnarray}
&&\gamma(s)\left[s\xdist(s)-\xdist(t=0)\right]
	=\noise(\na,s)-\noise(\nd,s) \nonumber \\
&& \ \ 	+\left(f_\fa(s)-f_\fd(s)\right)
	\left(1-e^{-\sqrt{s\kappa \xi}/ \kdist}\right),
\label{eq:FLEsubt}
\end{eqnarray}
in Laplace space where $\kdist = \kappa/|\na-\nd|$. Dividing Eq. (\ref{eq:FLEsubt}) by
$2(1-e^{-\sqrt{s\kappa\xi}/ \kdist})$ will result in the last term on the
right hand side becoming a force $\fdist=(f_\fa-f_\fd)/2$ which is conjugate
to the coordinate $\xdist$. Thus we rearrange Eq. (\ref{eq:FLEsubt})
to the form of the GLE:
\begin{equation}
\int_0^t d t'\; {\cal K}(t-t') \frac{d\xdist(t')}{d t'}=
\etadist(t)- {\cal U}'(\xdist) + \fdist(t),
\label{eq:distGLE}
\end{equation}
with friction kernel and a noise in Laplace space given by
\begin{eqnarray}
{\cal K} (s)&=&\frac{\gamma(s)}{2(1-e^{-\sqrt{s\kappa \xi}/ \kdist})}
-\frac \kdist s,\\
\etadist(s)&=& 
\frac{\noise(\na,s)-\noise(\nd,s)}{2(1-e^{-\sqrt{s\kappa\xi}/\kdist})}-
\frac{\kdist}{s} \delta\xdist(t=0), \nonumber 
\end{eqnarray}
where $\delta \xdist(t)=\xdist(t)-(\na-\nd)/\rho$. The constant terms
subtracted in the definitions of the friction and noise are combined in the
harmonic force $-{\cal U}'(\xdist)=-\kdist [\xdist(t)-(\na-\nd)/\rho]$. The
spring constant $\kdist$ corresponds exactly to the $|\na-\nd|$
$\kappa$-strength
springs that are connected in series in between the donor-acceptor particles
after harmonization.
A straightforward but lengthy calculation (see Appendix \ref{app:gle}) confirms that $\langle
\etadist(t)\etadist(t')\rangle=k_B T {\cal K}(|t-t'|)$, as required by the
fluctuation-dissipation theorem.
One can invert the kernel ${\cal K}(s)$
exactly and express it as a Jacobi theta function
\begin{equation}
{\cal K}(t)=
\frac{\gamma(t)}{4}\left[1+\vartheta_3\left(0,e^{-\tdist/4t}\right)\right]-\kdist ,\label{eq:K_t}
\end{equation}
where 
$\vartheta_3(z,u) = 1 + 2\sum_{m=1}^\infty u^{m^2}\cos\left(2 m z \right)$
and $\tdist = \kappa \xi /\kdist^2$. Note that $\tdist=(|\na-\nd|/\rho)^2/D_c$ so that $\tdist$ can be interpreted as the time it takes for the information
about the motion of particle $\na$ to diffuse to $\nd$. Examining 
Eq. (\ref{eq:K_t}) for $t \ll \tdist$ one finds 
${\cal K}(t) \sim \sqrt{\kappa \xi/(\pi t)}$ which implies that the two particles do not influence each other.
For longer times $t\gg \tdist$ we have 
${\cal K}(t) \sim \sqrt{\kappa \xi/(4\pi t)}$.
The factor $1/2$ difference in the prefactor of ${\cal K}(t)$ at long and short times means that the autocorrelation function will not decay exactly as a Mittag-Leffler function. Instead from Eq. (\ref{eq:distGLE}) we obtain 
\begin{equation}
\langle \delta \xdist(t)\delta \xdist(0)\rangle = 
\frac{k_BT}{\kdist} \left[
			{\rm erf} 	\sqrt{\frac{\tdist}{4 t}}
			-\frac{1-e^{-\tdist/4 t}}{\sqrt{\pi\tdist/4 t}}
	\right],
\label{eqDISTpred}
\end{equation}
with $\langle \delta \xdist(t)\delta \xdist(0)\rangle\sim k_B T \kdist^{-1}\sqrt{\tdist/(4\pi t)}$ at long times.
In Fig. \ref{fig1} we compare Eq. (\ref{eqDISTpred}) to
simulations and find excellent agreement (without fitting).

%
%

\section{Summary and concluding remarks}
Throughout this paper we have shown that our harmonization technique can reproduce known results as well as providing new ones. But how come it works so well? When equation (\ref{eq04}) was obtained, a quasi-static
approximation was used in the sense that the effective spring constants were
calculated based on the equilibrium properties of the system. To see why this
is physically reasonable one can argue as follows. The MSD of a tracer
particle in a single-file system is proportional to $t^{1/2}$ and it will therefore
cross a system of length $L$ in a time on the order of $L^4$. This is
considerably slower than the relaxation time of the whole system which scales as
$L^2/D_c$. Thus, in the long-time limit a tracer particle only sees particles
which have had sufficient time to reach local thermal equilibrium
\cite{note7}.
This is the reason why equilibrium concepts like free energy work so well here.
This implies that it is the one-dimensional topology and the single-file condition that leads to slow dynamics of a tracer particle and the possibility to map it onto a harmonic chain.  The FLE with exponent $1/2$ is therefore expected to be found in a vast number of over-damped systems. 
Our framework can, however, be applied to particle motion in higher dimensions. For instance, particles embedded in networks in which ordering is maintained.

In summary, we have presented a harmonization technique which maps a stochastic
many-particle system with general two-body potentials onto a system of interconnected springs. The interaction potential was reduced through equilibrium considerations to only one parameter: the spring constant $\kappa$ related to the compressibility of the particle system
\cite{note8}.
We derived, from first-principles, an FLE which predicted subdiffusive (slow) dynamics of a tracer particle. Derived expressions agree perfectly with rigorous well-known results when they are available. Under the influence of an external harmonic force, Mittag-Leffler relaxation was found which was corroborated by simulations of a hard-core system. It would be interesting to test
the harmonization
technique further with simulations beyond this hard sphere model. The dynamics of inter-particle distance was also addressed and the harmonization approach predicted a GLE rather than, as previously suggested, an FLE.
Unlike the ordinary Langevin equation, which describes a Markovian process and which is usually derived for a massive particle colliding with independent gas particles, the FLE exhibits long memories which, as we showed here, are due to the many-body nature of the underlying dynamics. Thus, fractional calculus enters through many-body effects which might be the reason why it took so long for a natural microscopic origin to be uncovered.

%
%

\section{Acknowledgements}
We acknowledge Bob Silbey, Igor Sokolov, 
Mehran Kardar, Aleksei Chechkin, John Ipsen and
Ralf Metzler for discussions. The work was supported by the Knut and
Alice Wallenberg foundation, the Israel Science Foundation and the Danish
National Research Foundation.

\appendix

\section{Mean square displacement of a tracer  particle}
\label{app:msd}

In this appendix we calculate the MSD of the tracer particle position $x_\mT(t)=x(n=0,t)$ defined as
\begin{equation}
\msd  \equiv \langle [x_\mT(t)-x_\mT(t=0)]^2\rangle,
\end{equation} 
under the assumption of no external force  $f(n,t)=0$. In order to find the MSD we calculate the correlation $\langle x(q,s)x(q',s') \rangle$ which we find by multiplying Eq. (\ref{eq10}) by itself and  average over the noise $\langle \ldots \rangle$:
\begin{eqnarray}
&& \langle x(q,s)x(q',s') =A_{\rm init}(q,q',s,s')+A_{\rm noise}(q,q',s,s'),\nonumber\\
&&A_{\rm init}(q,q',s,s')=\frac{\nu^2 \langle x(q,t=0)x(q',t=0)\rangle}  {(s\nu+q^2)(s'\nu+q'^2)},\\
&&A_{\rm noise}(q,q',s,s')=\frac{1}{\kappa^2} \frac{ \langle \eta(q,s) \eta(q',s') \rangle}  {(s\nu+q^2)(s'\nu+q'^2)},
\end{eqnarray}
where $\nu=\xi/\kappa$ and the variable name $q'$ ($s'$) like $q$ ($s$) implies that the corresponding variable have been Fourier (Laplace) transformed. Above we used that the initial positions are independent of the future thermal noise.

Starting with $A_{\rm noise}$ we first note that
the Fourier and Laplace transform of the noise autocorrelation function is
\begin{equation}
\langle \eta(q,s) \eta(q',s') \rangle = \frac{4\pi\xi k_BT \delta(q+q')}{s+s'}.
\end{equation}
Thus we can write
\begin{equation}
A_{\rm noise}(q,q',s,s')=  
\frac{4\pi\xi^{-1} k_BT \delta(q+q')}
{(s+s')(s+q^2/\nu)(s'+q^2/\nu)}.
\end{equation}
Taking the inverse Laplace transforms of this we find
\begin{equation}
A_{\rm noise}(q,q',t,t') =
2\pi \delta(q+q')
\frac{e^{-q^2|t-t'|/\nu} - e^{-q^2(t+t')/\nu} }{\kappa q^2/(k_B T)}.
\end{equation}
If we take the inverse Fourier transforms of the above equation and evaluate it $n=n'$ and $t=t'$ we find
\begin{eqnarray}\label{eq:msd_Anoise}
A_{\rm noise}(n,n,t,t)& =&
\frac{ k_BT } {\kappa} \int_{-\infty}^\infty \frac{dq}{2\pi}
\frac{1 - e^{-2q^2 t/\nu} }{q^2}\nonumber\\
& =& k_B T \sqrt{\frac{2t}{\pi \xi \kappa}}
\end{eqnarray}
where we used $\int_{-\infty}^\infty dz\, z^{-2}[1-e^{-az^2}] = 2\sqrt{a \pi}$.

We now proceed to evaluate $A_{\rm init}$. Using the inverse Fourier transform ${\cal F}_q^{-1}\{2a/(a^2+q^2)\} = e^{-a|n|}$ as well as the convolution theorem we can write
\begin{eqnarray}
A_{\rm init}(n,n',s,s') &=& 
\int_{-\infty}^{\infty} du \int_{-\infty}^{\infty} du'  
\frac {e^{-|n-u|\sqrt{s\nu}}}{2\sqrt{s/\nu}} 
\label{eq:Ainit}
\\
&\times&
\frac {e^{-|n'-u'|\sqrt{s'\nu}}}{2\sqrt{s'/\nu}}
\langle x(u,t=0)x(u',t=0)\rangle\nonumber
\end{eqnarray}
In the harmonic chain, the particles are in thermal equilibrium with respect to the potential
\begin{equation}\label{eq:harmPot}
U =\frac \kappa 2 \sum_m \, 
\left[x_m - x_{m-1}-\rho^{-1}\right]^2
\end{equation}
from which  the equilibrium density is ${\cal P}_{\rm equilib.} = e^{-U/(k_BT)}/Z$ where $Z = \int [\Pi_m dx_m]\, e^{-U/(k_BT)}$. Thus, the particles' initial positions are Gaussian variables which we can express as
\begin{eqnarray}
x_n(t=0) &=&\left\{
 \begin{array}{l l}
 	\sum_{r=1}^n \Delta_r &\;\; n>0   \\
						0 & \;\; n=0  \\
	-\sum_{r=1}^{|n|}\Delta_{-r} &\;\; n<0
 \end{array}\right.
\end{eqnarray}
where we have chosen the coordinates such that $x_0(t=0)=0$. The expected values of the $\Delta_r$ are
\begin{eqnarray}
&&\langle\Delta_r -\rho^{-1}\rangle= 0,\\ 
&&\langle(\Delta_r -\rho^{-1})(\Delta_{r'} -\rho^{-1})\rangle = \frac{k_BT}{\kappa} \delta_{r,r'}\,,
\end{eqnarray}
where $\delta_{r,r'}$ is the Kronecker delta.
From this we find
\begin{eqnarray}
&&\langle\left[x(u,t=0)-x(0,t=0)\right] \left[x(u',t=0)-x(0,t=0)\right]\rangle
\nonumber\\
&& = u u' \rho^{-2}+{\rm min}(|u|,|u'|)\theta(u u')\frac{k_B T}{\kappa},\label{eq:initcor}
\end{eqnarray}
where $\theta(x)$ is the Heaviside step function and
${\rm min}(a,b)$ is the smallest value of $a$ and $b$.
Inserting this initial distribution in Eq. (\ref{eq:Ainit}) and setting $n=n'=0$ leads to
\begin{eqnarray}
&&A_{\rm init}(n=0,n'=0,s,s') \nonumber\\ 
&&= \frac{ k_BT}{\kappa} 
\int_{0}^{\infty} du \int_{0}^{\infty} du'   
\,2\, {\rm min} (u,u')
\frac {e^{-u\sqrt{s\nu}}}{2\sqrt{s/\nu}} 
\frac {e^{-u'\sqrt{s'\nu}}}{2\sqrt{s'/\nu}}
\nonumber \\
&&=  \frac{\nu^2k_BT}{2 \kappa} 
\int_{0}^{\infty} du \int_{0}^{u} du'   
u' \frac {e^{-u\sqrt{s\nu}}}{\sqrt{s\nu}} 
\frac {e^{-u'\sqrt{s'\nu}}}{\sqrt{s'\nu}}\nonumber\\
&&\;\;\;\;+ \frac{ \nu^2k_BT}{2\kappa} 
\int_{0}^{\infty} du' \int_{0}^{u'} du\,   
u \frac {e^{-u\sqrt{s\nu}}}{\sqrt{s\nu}} 
\frac {e^{-u'\sqrt{s'\nu}}}{\sqrt{s'\nu}}.
\end{eqnarray}
Using the inverse Laplace transform 
\begin{equation}
{\cal L}^{-1} \left\{ \frac {e^{-u\sqrt{s\nu}}}{\sqrt{s\nu}} \right\} =
\frac{e^{-u^2\nu/4t}}{\sqrt{\nu\pi t}},
\end{equation}
evaluating the second integral in each of the terms above, and setting $t=t'$ gives
\begin{eqnarray}
&&A_{\rm init}(n=0,n=0,t,t)\nonumber\\
 &&= \frac{2}{\pi} \frac{ k_B T}{\kappa} 
\int_{0}^{\infty} du
\left(
 e^{-u^2\nu/4t}-e^{-2u^2\nu/4t}
\right) \nonumber\\
&&= 
k_BT \sqrt{\frac{t}{\pi\xi\kappa}} (2-\sqrt{2}),
\end{eqnarray}
where $\nu = \xi/\kappa$ was used. If $A_{\rm init}$ is combined with Eq. (\ref{eq:msd_Anoise}) we find the desired result for the MSD for a tracer particle in thermal equilibrium
\begin{eqnarray}
\msd &=& A_{\rm init}(n=0,n=0,t,t)+A_{\rm noise}(n,n,t,t)\nonumber\\
&=&k_BT \sqrt{4t/(\pi\xi\kappa)},
\end{eqnarray}
which is the result mentioned in the main text. We note that if the particles initially had been placed equidistantly, $x(n,t=0)=n/\rho$, with no randomness in the positioning, then $A_{\rm init}$ would have vanished and the MSD would have been smaller by a factor $\sqrt{2}$.

\section{Fluctuation-dissipation relation for $\bar{\eta}^{\rm eff}$}
\label{app:fle}
Here we will find the noise-correlation in Laplace space for the more general case $\left<\bar{\eta}^{\rm eff}(n,s)\bar{\eta}^{\rm eff}(n',s')\right>$ from which the fluctuation-dissipation relation follows as the special case $n=n'$. The more general case will be needed in Appendix \ref{app:gle}.

We will divide the noise into two parts
\begin{equation}
\bar{\eta}^{\rm eff}(n,s)=\bar{\eta}(n,s)+\etainit(n,s),
\end{equation}
where the first part is related to the original thermal noise
\begin{equation}
\bar{\eta}(n,s)=\int_{-\infty}^\infty d u\, e^{-\sqrt{s\nu}|n-u|}\eta(u,s),
\end{equation}
and the second part is related to the initial positions
\begin{eqnarray}
\etainit(n,s)&=&\int_{-\infty}^\infty d u\,e^{-\sqrt{s\nu}|n-u|}\nonumber\\
&&\;\;\times\xi\left[x(u,t=0)-x(n,t=0)\right].
\end{eqnarray}
For the first part we have the correlation function
\begin{eqnarray}
&&\left<\bar{\eta}(n,s)\bar{\eta}(n',s')\right>=
\int_{-\infty}^\infty d u\,e^{-\sqrt{s\nu}|n-u|}
\nonumber\\
&&\times\int_{-\infty}^\infty d u'\, e^{-\sqrt{s'\nu}|n'-u'|}\, \frac{2\xi k_B T}{s+s'}\delta(u-u').
\end{eqnarray}
Doing the integrals one finds
\begin{eqnarray}
&&\left<\bar{\eta}(n,s)\bar{\eta}(n',s')\right>=
\frac{4 \kappa k_B T}{(s+s')(s-s')}
\nonumber\\
&&\;\;\;\times\left(\sqrt{s\nu}\, e^{-\sqrt{s' \nu}|n-n'|}-\sqrt{s'\nu}\, e^{-\sqrt{s\nu}|n-n'|}\right).
\end{eqnarray}
For the part of the noise correlation that comes from the random initial condition we have
\begin{eqnarray}
&&\left<\etainit(n,s)\etainit(n',s')\right>=\xi^2 \int_{-\infty}^\infty d u\,e^{-\sqrt{s\nu}|n-u|}
\nonumber\\
&&\times \int_{-\infty}^\infty d u'\, e^{-\sqrt{s'\nu}|n'-u'|}\, \left<\left[x(u,t=0)-x(n,t=0)\right]\right.
\nonumber\\
&&\times\left. \left[x(u',t=0)-x(n',t=0)\right]\right>
\end{eqnarray}
Using Eq. (\ref{eq:initcor}) one finds after a bit of calculation that
\begin{eqnarray}
\langle\etainit(n,s)\etainit(n',s')\rangle&=&\frac{2 k_B T \xi}{s-s'}\left(-\frac{1}{\sqrt{s\nu}}\, e^{-\sqrt{s' \nu}|n-n'|}\right. 
\nonumber\\
&&\left.+\frac{1}{\sqrt{s'\nu}}\, e^{-\sqrt{s\nu}|n-n'|}\right).
\end{eqnarray}
Combining the two noise parts leads to the general formula
\begin{eqnarray}
&&\left<\bar{\eta}^{\rm eff}(n,s)\bar{\eta}^{\rm eff}(n',s')\right>=
\nonumber\\
&&\frac{2 k_B T \sqrt{\xi \kappa}}{s+s'}\left(\frac{1}{\sqrt{s}}\, e^{-\sqrt{s' \nu}|n-n'|}+\frac{1}{\sqrt{s'}}\, e^{-\sqrt{s\nu}|n-n'|}\right)=\nonumber\\
&&\frac{k_B T}{s+s'}\left(\gamma(s)\, e^{-\sqrt{s' \nu}|n-n'|}+\gamma(s')\, e^{-\sqrt{s\nu}|n-n'|}\right).\label{eq:genetaeff}
\end{eqnarray}
Setting $n=n'$ finally gives
\begin{equation}
\left<\bar{\eta}^{\rm eff}(n,s)\bar{\eta}^{\rm eff}(n,s')\right>=\frac{k_B T\left[\gamma(s)+\gamma(s')\right]}{s+s'}
\end{equation}
which is the Laplace transform of the sought for fluctuation-dissipation relation $\langle \bar{\eta}^{\rm eff}(n,t) \bar{\eta}^{\rm eff}(n,t')\rangle = k_BT\gamma(|t-t'|)$.


\section{External force on a tracer  particle}
\label{app:force}
Here we consider a force acting only on particle 0:
\begin{equation}\label{eq:F}
f(n,t) = F(t)  \delta(n).
\end{equation}
Taking the average of Eq. (\ref{eq10}) with respect to the zero-mean noise and using the inverse Fourier transform ${\cal F}_q^{-1}\{2a/(a^2+q^2)\} = e^{-a|n|}$ with the explicit expression for the force Eq. (\ref{eq:F}) leads to
\begin{eqnarray}\label{eq:x_avg2}
&&\langle x(n,s) \rangle =
\nonumber\\
&&\nu \int_{-\infty}^{\infty} du \, \langle x(u,t=0)  \rangle
\frac{e^{-\sqrt{s\nu}|n-u|}}{2\sqrt{s\nu}}
+ \frac{F(s)} {2\kappa}  \frac{e^{-\sqrt{s\nu}|n|}}{\sqrt{s\nu}}
\nonumber\\
&&=\frac{n\rho}{s}+ \frac{F(s)} {2\kappa}  \frac{e^{-\sqrt{s\nu}|n|}}{\sqrt{s\nu}}.
\end{eqnarray}
Here we used that $\langle x(n,t=0)\rangle = n/\rho$.

\subsection{Periodic force $f(n,t) = \delta(n) F_0\cos (\omega_0 t) $ }
\label{sec:FL_Force_periodic}
If we use the complex representation of the force $f(n,t) = \delta(n) \Re [F_0 e^{i\omega_0 t}]$ (where $\Re$ represents the real part) we have the Laplace transform
\begin{equation}
f(n,s) = \delta(n) \Re\left[\frac{F_0}{s-i\omega_0}\right].
\end{equation}
Using this in Eq. (\ref{eq:x_avg2}) gives
\begin{equation}
\langle x(n,s) - \overbrace{x(n,t=0)}^{=n/\rho} \rangle = 
\Re \left[
 \frac{F_0} {2\kappa}  \frac{e^{-\sqrt{s\nu}|n|}}{\sqrt{s\nu}(s-i\omega_0)}
\right].
\end{equation}
For $n=0$, i.e we track the tagged particle on which the force acts, we have
\begin{equation}
\langle x_\mT(t) \rangle = 
\Re \left[
 \frac{F_0} {2\kappa}  \frac{e^{i\omega_0 t} \, {\rm Erf}(\sqrt{i\omega_0t})}{\sqrt{i\omega_0\nu}}
 \right],
\end{equation}
where we used ${\cal L}^{-1}\left\{ (\sqrt{s}(s-a))^{-1} \right \}= e^{at}\,{\rm Erf}(\sqrt{at})/\sqrt a$. Making the replacement $\nu=\xi/\kappa$ and $i^{1/2}=e^{i\pi/4}$ leads to
\begin{equation}\label{eq:x_avg_cos}
\langle x_\mT(t) \rangle = 
 \Re \left[
 \frac{F_0} {2\sqrt{\omega_0\kappa \xi}}  e^{i(\omega_0 t-\pi/4)} {\rm Erf}(\sqrt{i\omega_0t})
 \right].
\end{equation}
For large $t$ we have
${\rm Erf}(\sqrt{i\omega_0t}) \sim 1$
and therefore
\begin{equation}
\langle x_\mT(t) \rangle \sim
 \frac{F_0} {2\sqrt{\omega_0\kappa \xi}} \cos\left( \omega_0 t-\pi/4\right)
\end{equation}
after taking the real part.

\subsection{Constant force $f(n,t) =\delta(n) F_0 $ }

The result for a constant force is obtained in the $\omega_0 \rightarrow 0$ limit
of Eq. (\ref{eq:x_avg_cos}). Using the short time expansion 
\begin{equation}
{\rm Erf}(\sqrt{i\omega_0t}) \sim 2\sqrt{\frac{i\omega_0t}\pi}
= 2\sqrt{\frac{\omega_0t}\pi}e^{i\pi/4} 
\end{equation}
combined with Eq. (\ref{eq:x_avg_cos}) gives
\begin{equation}\label{eq:x_avg_const}
\langle x_\mT(t) \rangle = 
 F_{0}\sqrt{\frac{t}{\pi \kappa \xi}}.
\end{equation}
Together with Eq. (\ref{eq05}) this demonstrates the generalized Einstein relation, Eq. (\ref{eq:einstrel}) in the main text.

\section{Fluctuation-dissipation relation for $\etadist$}
\label{app:gle}
Here we address the fluctuation-dissipation relation for $\etadist$. Most of the work for deriving this was done in Appendix \ref{app:fle} when deriving Eq. (\ref{eq:genetaeff}). What remains is to work out the correlations of the part $\etadist^{\rm init}(s)\equiv \kdist\delta\xdist(t=0)/s$. These turn out to be:
\begin{eqnarray}
&&\left<\etadist^{\rm init}(s)\etadist^{\rm init}(s')\right>=\frac{k_B T \kdist}{s s'},
\nonumber \\
&&\left<\noise({\na},s)\etadist^{\rm init}(s')\right>=\frac{k_B T \kdist}{s s'}\left(1-e^{-\sqrt{s\nu}|\nd-\na|}\right),
\nonumber \\
&&\left<\noise({\nd},s)\etadist^{\rm init}(s')\right>=-\left<\noise({\na},s)\etadist^{\rm init}(s')\right>\;.\nonumber
\end{eqnarray}
Combining these correlations with Eq. (\ref{eq:genetaeff}) one arrives at
\begin{equation}
\langle\etadist(s)\etadist(s')\rangle=\frac{k_B T}{s+s'}\left[{\cal K}(s)+{\cal K}(s')\right],
\end{equation}
which is the Laplace transform of the sought for relation: $\langle
\etadist(t)\etadist(t')\rangle=k_B T {\cal K}(|t-t'|)$.

%
%

\bibliography{references}

\begin{thebibliography}{48}
\bibitem[{not({\natexlab{b}})}]{note2}
\bibinfo{note}{Long-ranged interactions yielding non-extensive behavior, e.g.
  an unscreened Coulomb potential, are omitted.}
\bibitem[{not({\natexlab{e}})}]{note5}
\bibinfo{note}{We point out that the case with thermal (equilibrium) initial
  conditions corresponds to using a Fourier transform in time. However, our
  Laplace analysis in time domain is more general and allows studies of other
  types of initial distributions.}
\bibitem[{not({\natexlab{f}})}]{note6}
\bibinfo{note}{The damping term, i.e., the left hand side of Eq. (\ref{eqFLE}),
  can be written as $\int_0^t dt'\,\gamma(t-t')d x_\mT(t')/dt'$.}
\bibitem[{not({\natexlab{g}})}]{note7}
\bibinfo{note}{For hard core-particles in higher dimensions the quasi-static
  approximation is expected to break down, at least for low particle densities
  since a tracer particle diffuses normally invalidating the above argument.}
\bibitem[{not({\natexlab{h}})}]{note8}
\bibinfo{note}{The compressibility of the particle system does not necessarily
  involve the compressibility of the medium between the particles. For
  instance, if the particles are embedded in water that can flow past the
  particles, then the statistical mechanical calculation of the compressibility
  of the particle system should involve water kept at a uniform chemical
  potential.}
\end{thebibliography}

\end{document}